\documentclass[showpacs,amsmath,amssymb,twocolumn,floatfix,prl]{revtex4}

\newcommand{\arctanh}{\mbox{arctanh}}
\usepackage{graphicx}
\input{epsf}
\begin{document}

\title{Photocurrent in nanostructures with asymmetric antidots}
\author{M.V. Entin,  L.I. Magarill}
\affiliation{Institute of Semiconductor Physics, Siberian Branch
of Russian Academy of Sciences, Novosibirsk, 630090, Russia}

\begin{abstract}
The steady current induced by electromagnetic field in a 2D system
with asymmetric scatterers is studied. The scatterers are assumed
to be oriented cuts with one diffusive and another specular sides.
Besides, the existence of isotropic impurity scatterers is
assumed. This simple model simulates the lattice of half-disk
which have been studied numerically recently. The model allows the
exact solution in the framework of the kinetic equation. The
static current response in the second order of electric field is
obtained. The photogalvanic tensor contains both responses to
linear and circular polarization of electromagnetic field. The
model possesses  non-analyticity with regards to the rate of
impurity scattering.
\end{abstract}

\maketitle

\section{Introduction}
Nowadays  technology allows to fabricate artificial arrays of
antidot scatterers which form superlattices in semiconductor
heterostructures with a two dimensional electron gas (2DES) (see
e.g. \cite{weiss,kvon,kvon1,weiss1} and Refs. therein). The size
of antidots can be varied from a few microns to a few tens of
nanometers at a typical electron density $n_e \sim
10^{12}cm^{-2}$. Superlattices with circular antidots (disks),
known as the Galton board \cite{galton}, have been realized in
experiments  \cite{weiss,kvon,kvon1,weiss1}. The mathematical
theorems of Sinai garanty that the classical dynamics of electrons
in such structures is chaotic \cite{sinai}. The effects of chaotic
dynamics and contributions of unstable periodic orbits on electron
conductivity have been clearly seen experimentally. They were also
analyzed by theoretical methods and numerical simulations in great
detail \cite{geisel}. The irradiation of such superlattices by a
microwave field \cite{kvon2}  opens interesting possibilities for
microwave control of electron current in nanostructures. These
systems of relatively large period are classical, and the periodic
potential is the source of electron scattering.

Unlike arrays of symmetric  antidots,  studied experimentally in
early works, more sophisticated systems were for a long time out
of attention. Meanwhile,  systems without inversion symmetry are
capable to rectify the electric current. The stationary current in
homogeneous media affected by light in the absence of any {\it
dc-}voltage called the {\it photogalvanic effect} (PGE) was
studied since the end of 70th
\cite{bbem,bem,bel,belin,ivch,pikus}. Similar direct current
caused by temporal irreversibility due to simultaneous action of
two electromagnetic fields with frequencies $\omega$ and $2\omega$
is known as the coherent photogalvanic effect (CPGE)) \cite{ent}.

Appearance of directed transport without obvious directed forces
is also known as ratchet effect which has a long history. For
example,  the behavior of a ratchet  under the influence of
thermal fluctuations was considered  in the textbook by Feynman,
Leighton and Sands \cite{feyn} in connection with the problem of
reversibility in statistical mechanics. Recently the ratchet
problem attracted a great interest of the scientific community
\cite{hanggi,reimann}. In fact the photogalvanic effect
corresponds to a ratchet subjected to weak alternating force with
zero mean. Such ratchets have been observed in various physical
systems including Josephson junction arrays
\cite{mooij,nori,ustinov}, cold atoms \cite{grynberg}, macroporous
silicon membranes \cite{muller}, microfluidic channels
\cite{ajdari} and other systems. The growing interest to ratchets
is strongly stimulated by their possible applications to
biological systems \cite{hanggi,ajdari2}. In this sense the
artificial asymmetric nanostructures, as those discussed in
\cite{semidisk1,cristadoro,semidisk2} and here, can serve as a
prototype for understanding of photocurrent properties in
biomolecules. The results obtained for ratchets induced by
microwave fields in nanostructures can be also used for
understanding of directed transport created by {\it ac-}fields in
molecular electronics \cite{ratner}.

Relatively recently an artificial  lattice of asymmetric antidots
(triangles) participating in transport as scatterers of electrons
was realized experimentally. It exhibited the direct current
induced by alternating electric field \cite{lorke}. The effect of
such type  was considered theoretically in \cite{magar} by means
of the classical kinetic equation for the system with weak
asymmetric periodic lateral
 potential.

Another case of superlattices of asymmetric antidots (semidisks
oriented in one direction or the semidisk Galton board) has been
proposed and analyzed theoretically
\cite{semidisk1,cristadoro,semidisk2} by simulations of motion of
a particle subjected to alternating force with zero means and
collisions with hard-wall antidots. It has been shown that a
microwave radiation  creates the directed flow of electrons. The
velocity of this flow $v_f$ is proportional to a friction
coefficient  while the direction of current depends on the
microwave polarization. The directed transport induced by a
microwave field appears also for an ensemble of noninteracting
particles been at a thermal equilibrium at temperature $T$.  This
effect is absent in structures with circular antidots due to
symmetry conservation. Recently, semidisk Galton board realized
experimentally \cite{kvon3} exhibited the rectification of
high-frequency electric field.

In \cite{semidisk1,cristadoro,semidisk2} the properties of
photocurrent in asymmetric nanostructures have been explained on
the basis of heuristic arguments and extensive numerical
simulations. However, the analytical theory of the effect still
needs to be developed. This aim is reached in this paper with the
help of a kinetic equation approach applied to a specific model of
asymmetric scatterers (cuts model). This approach allows to obtain
analytical dependence of photocurrent properties on system
parameters.

\section{Kinetic equation approach}

Here we study a simple model of anisotropic 2D artificial
scatterers, which permits the analytical
 consideration in the framework of kinetic equation
 approximation and leads to the PGE. The kinetic equation was used in a number of papers devoted to
photogalvanic effect. Unlike the particle dynamics method, this
way  suggests the developed chaos picture where ergodicity  of
motion is achieved. At the same time it is free from the voluntary
assumptions
 about the particle friction used in \cite{semidisk1,cristadoro}.

\begin{figure}[ht]
\centerline{\epsfysize=4cm \epsfbox{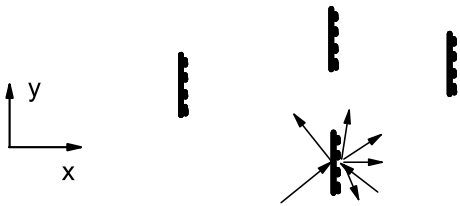}}
\caption{Considered model system. Cuts have specular left sides
and diffusive right sides. This produces anisotropy of scattering
resulting in the photocurrent. } \label{cuts}
 \end{figure}

The  system under consideration contains randomly distributed
oriented scatterers. The scatterers are assumed to be segments of
the length $D$ oriented along the $y$ axis; one side (left) of the
segment is specular and the other side is diffusive (see Fig.
\ref{cuts}).
 The concentration of scatterers  $N$ is supposedly low:
$ND^2\ll 1$.
 In this approximation the kind of spatial distribution of scatterers
  (random or periodic) is of no importance.  Besides, to limit the possible
  divergency of the result we include  isotropic impurity scattering into our model.

This system can be considered as a simplification of the
semicircle model studied in \cite{semidisk1,cristadoro}. In fact,
the diffusive side of the cut scatters particles like round side
of the semicircle, if not to pay attention on the difference
between randomized (in our case) and deterministic (in semicircle
case) motion.  The advantage of "cuts" model is its exact
solvability.

 The kinetic equation reads as
   \begin{equation}\label{kin}
\frac{\partial f}{\partial t} + \hat{F}f= \hat{I}f,
\end{equation}
where  $f(p,\varphi)$ is the distribution function,  ${\bf
p}=p(\cos{\varphi},\sin{\varphi})$ is the electron momentum.
 The term ($\hat{F}f$) represents the action of electric field
 ${\bf E}(t) =\mbox{Re}({\bf E}_\omega e^{-i\omega t})$  of the
 electromagnetic wave with the complex amplitude ${\bf E}_\omega={\bf E}_{-\omega}^*$:
 \begin{eqnarray}\label{F}
  && \hat{F}\equiv\frac{1}{2}\hat{F}_\omega e^{i\omega t}+c.c.=-e[E_x(\cos{\varphi}\frac{\partial}{\partial p}-
\frac{\sin{\varphi}}{p}\frac{\partial}{\partial \varphi} )+
\nonumber\\&& +E_y(\sin{\varphi}\frac{\partial}{\partial p}+
\frac{\cos{\varphi}}{p}\frac{\partial}{\partial \varphi} )]
\end{eqnarray}
As it will be seen further, the acceleration in the $y$ direction
can not be limited by the scattering on segments only  because
electrons moving in this direction do not relax. This is why we
have not restricted our consideration by the collision with cuts
only but have also taken into account the impurity scattering.
Hence the collision integral $\hat{I}f\equiv
\hat{I}_if+\hat{I}_cf$ is assumed to consist of impurity-induced
$\hat{I}_i$ and segment-induced $\hat{I}_c$ collision integrals.
The segment-induced collision integral
\begin{eqnarray}\label{IF}
 \hat{I}_cf = \int_0^{2\pi}
d\varphi'[W(\varphi',\varphi)f(p,\varphi')-W(\varphi,\varphi')f(p,\varphi)].
\end{eqnarray}
 is determined by  the scattering probability on the cuts
 $W(\varphi',\varphi)$:
\begin{widetext}\begin{equation}\label{W}
    W(\varphi',\varphi)=\frac{1}{\tau} \left[\cos \varphi'~\theta(\cos \varphi')\delta(\varphi'+\varphi-\pi)-
    \frac{1}{2}\cos \varphi' \cos \varphi~\theta(\cos \varphi)\theta(-\cos
    \varphi')\right],
\end{equation}\end{widetext}
where  $\tau=(DNv)^{-1}$ is the corresponding characteristic time,
$ v=p/m$; $\theta(x)$ is the Heaviside function. The impurity
scattering is suggested to be isotropic:
\begin{eqnarray}\label{IiF}
\hat{I}_if =-\frac{1}{\tau_{i}}(f-<f>),
\end{eqnarray}
where $<...>$ means average over angles, $\tau_i$ is the
relaxation time corresponding additional isotropic scattering.

 We shall solve the kinetic equation Eq.(\ref{kin})in the second order on electric
field. In the elastic approximation the collision operator  is
degenerate because its action on the isotropic function gives
zero; thus this operator has  an eigenfunction with zero
eigenvalue. Physically, this means that the elastic scattering
does not change any distribution function depending only on the
energy. But if to project the Gilbert space of distribution
functions on the subspace with zero angular average, the
corresponding projection of the scattering operator becomes
non-degenerate. So if to consider only the distribution functions
with zero mean, the elastic scattering operator remains
non-degenerate and can be treated as full relaxation operator.
Inelastic scattering controls only isotropic part of the
distribution function which (if it is weaker than elastic one) has
no impact on the PGE.

In the second order on alternating electric field the steady
current density  is described  by the phenomenological expression
\begin{equation}\label{j}
    j_i=\alpha_{ijk} E_{\omega,j}E_{-\omega,k}.
\end{equation}
 The symmetry of the considered system
allows the following non-zero components of  the photogalvanic
tensor $\alpha_{ijk}$:
 \begin{equation}\label{al}
    \alpha_{xxx}, ~~\alpha_{xyy},~~\alpha_{yxy}=\alpha_{yyx}^*.
\end{equation}
From  Eq.(\ref{j}) it follows that components $\alpha_{xxx}$ and
$\alpha_{xyy}$ are real. The same is valid for all components of
$\alpha_{ijk}$ in the static limit ($\omega=0$).

The equation  (\ref{j})  is specified as
\begin{eqnarray} \label{jj}
 j_x =\alpha_{xxx}|E_x|^2+ \alpha_{xyy}|E_y|^2,\nonumber\\
 j_y= \mbox{Re}(\alpha_{yxy})(E_xE_y^*+ E_x^*E_y)+
 \mbox{Im}(\alpha_{yxy})[{\bf E} {\bf E}^*]_z.
\end{eqnarray}
The components $\alpha_{xxx}$, $\alpha_{xyy}$   and
$\mbox{Re}(\alpha_{yxy})$ determine the response to the
linear-polarized light. For  linear polarization along $x$ or $y$
axes the current flows along the $x$ direction; the current in $y$
direction appears for tilted linear-polarized electric field. In
the case of circular polarization the $y$ component of the current
is determined by $\mbox{Im}(\alpha_{yxy})$ (the circular
photogalvanic effect) and by the sign of the rotation, while $x$
component of the current is the sum of responses to $x$ and $y$
linear polarized light and  does not depend on the sign of $[{\bf
E} {\bf E}^*]_z$.  Notice that the circular photogalvanic effect
vanishes if the frequency $\omega\to 0$.

 The formal solution of the Eq.(\ref{kin}) in the second order in
electric field  is
\begin{equation}\label{sol}
    f_2=(\frac{\partial}{\partial t}-\hat{I})^{-1} \hat{F}(\frac{\partial}{\partial t}-\hat{I})^{-1}
    \hat{F}f_0
\end{equation}
where  $f_0=1/(\exp{(\epsilon-\mu)/T})^{-1}$ is the isotropic
equilibrium distribution function, $\mu$ is the chemical
potential, $T$ is the temperature. The kernel of the inverse
operator in frequency representation, the Green function
$G^\omega(\varphi',\varphi)$, satisfies  the equation
$(-i\omega+\hat{I})G_\omega(\varphi,\varphi')=\delta(\varphi-\varphi')$.
This equation can be solved exactly:
\begin{widetext}\begin{eqnarray}\label{G}
\nonumber
    G^\omega(\varphi,\varphi')=-\tau
    \Big\{\frac{\delta(\varphi-\varphi')}{|\cos\varphi|+\xi}+
    \frac{|\cos\varphi|}{2}\frac{\cos^2\varphi'-\xi\cos\varphi'~\theta(-\cos\varphi')}{(|\cos\varphi|+\xi)
    (|\cos\varphi'|+\xi)^2(1-\Phi(\xi))}+ \\
    \frac{\theta(-\cos\varphi)}{(|\cos\varphi|+
    \xi)^2}\left[-\cos\varphi~\delta(\varphi+\varphi'-\pi)+\xi\cos\varphi
    \frac{\cos^2\varphi'-\xi\theta(-\cos\varphi')\cos\varphi'}{2(|\cos\varphi'|+
    \xi)^2(1-\Phi(\xi))}\right]\Big\}
\end{eqnarray}\end{widetext}
The operator $G^\omega$ depends on the frequency via the parameter
$\xi=-i\omega\tau+ \tau/\tau_i$,
\begin{equation}\label{Phi}
    \Phi(x)=-\pi x +2-\frac{1}{1-x^2}
    -x^2\frac{3-2x^2}{1-x^2}\Phi(1,x).
\end{equation}
 The functions $ \Phi(n,\xi)$ denote the
integrals expressing via elementary functions:
\begin{eqnarray}\label{Phi1}
    \Phi(n,x) =\int_0^{\pi/2}\frac{ dt}{
    (\cos t+x)^n}, \nonumber \\ \Phi(1,x) =-\frac{2}{{\sqrt{1 - {x }^2}}} ~\arctanh(\frac{-1 + x }
      {{\sqrt{1 - {x }^2}}}).
\end{eqnarray}

 The stationary part of the correction
$f_2$ can be written as\begin{equation}\label{f2}
    \bar{f}_2= \frac{1}{4}G^0 (\hat{F}_\omega G^{-\omega}
    \hat{F}_{-\omega}+ \hat{F}_{-\omega} G^{\omega}
    \hat{F}_{\omega})f_0;  ~~~~~G^0 \equiv G^{\omega=0}.
\end{equation}
The static current reads \begin{equation}\label{j0}
    j_i=-2e\int\frac{d^2p}{(2\pi)^2} v_i\bar{f}_2.
\end{equation}
The photogalvanic tensor is expressed via partial tensor
$\tilde{\alpha}_{ijk}$ for electrons with given $p$ :
\begin{equation}\label{pge}
    \alpha_{ijk}=  \frac{e^3}{2m\pi^2}\int_0^\infty dp ~\tau^2
    (-\epsilon\frac{\partial f_0}{\partial \epsilon})
    \tilde{\alpha}_{ijk}.
\end{equation}
In the degenerate case  the partial tensor itself   determines the
total tensor $ \alpha_{ijk}$:
 \begin{equation}\label{al2}
    \alpha_{ijk}(T=0,\epsilon_F)=
    \alpha_0~\tilde{\alpha}_{ijk}|_{p=p_F},
\end{equation}
where $\alpha_0=e^3/(4\pi^2N^2D^2v_F), v_F=p_F/m, p_F$ is the
Fermi momentum, $\epsilon_F=p_F^2/2m$  is the Fermi energy. At
finite temperatures $T$
     \begin{equation}\label{al3}
    \alpha_{ijk}(T,\mu)= \int d\epsilon ~(-\frac{\partial f_0}{\partial \epsilon})
    ~ \alpha_{ijk}(T=0,\epsilon_F=\epsilon).
\end{equation}
 The dimensionless tensor   $\tilde{\alpha}_{ijk}$ can be
presented as a function of  parameters $\zeta=\tau/\tau_i$ and
$\xi=\zeta-i\Omega ~~(\Omega=\omega\tau)$.

\section{Analytical results}

Substituting the stationary part of distribution function
(\ref{f2})  into the expression (\ref{j0}) and using the Green
function (\ref{G}) we find after integration:
\begin{widetext}
\begin{eqnarray}\label{alxxx}\nonumber &&\tilde{\alpha}_{xxx}(\xi,\zeta)=\left( B(\zeta ) - B(\xi ) \right)  \left( 1 - \frac{\pi \,\zeta }{2} +
     \frac{ {\zeta }^4 + \zeta \xi  - 2{\zeta }^3\xi   }
      {{\left(\xi -\zeta  \right) }^2}\,\Phi (1,\zeta ) -
     \frac{ \zeta \xi    - 2\,\zeta \,{\xi }^3 +{\xi }^4  }{{\left(  \xi-\zeta   \right) }^2}
     \,   \Phi (1,\xi ) + {\xi }^2 \frac{{\xi }^2  -1  }{ \xi-\zeta   } \,\Phi (2,\xi )\right)  +
     \\ \nonumber
  &&B(\zeta )\,B(\xi )\Bigg( -\frac{ \pi \zeta   }{2} -
     \frac{2\,\zeta \,\left( {\zeta }^4 - \zeta \,\xi  - 3\,{\zeta }^3\,\xi  - {\xi }^2 +
          4\,{\zeta }^2\,{\xi }^2 \right) \,}{{\left(\xi -\zeta    \right) }^3}\Phi (1,\zeta ) +
     \frac{2\,\xi \,\left( {\xi }^4 + {\zeta }^2\,\left( -1 + 4\,{\xi }^2 \right)  -
          \zeta \,\left( \xi  + 3\,{\xi }^3 \right)  \right) }{{\left(
           \xi-\zeta  \right) }^3}\,\Phi (1,\xi ) - \\  &&\frac{{\zeta }^2\,
        \left( {\zeta }^3 + \xi  - 2\,{\zeta }^2\,\xi  \right) }{{\left( \xi-\zeta   \right) }^2}\Phi (2,\zeta ) + \frac{{\xi }^2\,
        \left( \xi  - 4\,{\xi }^3 + \zeta  \left( -4 + 7 {\xi }^2 \right)  \right) }
        {{\left( \xi-\zeta   \right) }^2} \Phi (2,\xi )  +
     \frac{2 {\xi }^3 \left( -1 + {\xi }^2 \right)}{ \xi-\zeta  }   \Phi (3,\xi
     )\Bigg)+c.c.
\end{eqnarray}

\begin{eqnarray}  \label{alxyy}
 && \tilde{\alpha}_{xyy}=1 - 3 B(\zeta ) + \frac{\pi\zeta
     \left(  4 B(\zeta ) -1  \right) }{2} +
   \left( 3 {\zeta }^3 \xi -{\zeta }^4 - {\zeta } \xi +  {\xi }^2 -
       2 {\zeta }^2 {\xi }^2 +B(\zeta )(5 {\zeta }^4  +
       3 {\zeta } \xi   -
       15 {\zeta }^3 \xi   -
       3 {\xi }^2 +
       10 {\zeta }^2 {\xi }^2 ) \right)
     \frac{\zeta\Phi (1,\zeta )}{{\left(\xi -\zeta  \right) }^3}\nonumber\\
    &&
   + \frac{\xi\left( {\zeta }^2   - \zeta  {\xi } -
       2 {\zeta }^2 {\xi }^2 + 3 \zeta  {\xi }^3 -
       {\xi }^4 + B(\zeta )( 3 \zeta  {\xi } - 3 {\zeta }^2 +
       10 {\zeta }^2 {\xi }^2  -
       15 \zeta  {\xi }^3  +
       5 {\xi }^4) \right)  \Phi (1,\xi )}
     {{\left(\xi -\zeta  \right) }^3}+\frac{\zeta^2B(\zeta )\left( {\zeta }^3  +
        \xi   -
       2 {\zeta }^2 \xi \right)
     \Phi (2,\zeta )}{{\left(\xi -\zeta   \right) }^2}
    +  \nonumber\\ &&\frac{\xi^2\left( \zeta   - {\xi } -
       \zeta  {\xi }^2 + {\xi }^3 +B(\zeta )(4 {\xi }-
       5 \zeta            +
       8 \zeta  {\xi }^2  -
       7 {\xi }^3 ) \right)  \Phi (2,\xi )}
     {{\left(\xi -\zeta   \right) }^2}
   +
  \frac{2 \xi^3B(\zeta )\left(
       {\xi }^2 -1) \right)  \Phi (3,\xi )}{
       \xi -\zeta }
\end{eqnarray}
 \begin{eqnarray}  \label{alyxy}
 &&\tilde{\alpha}_{yxy}(\xi,\zeta) =\Bigg( ( \zeta  - \xi   )
      ( -2\zeta \xi  +
       {\pi }^2\zeta
         ( \zeta  - \xi   )
         ( -1 + {\xi }^2  )  +
       \pi  ( -\xi  +
          \zeta  ( 2 +
              ( \zeta  - 3\xi   ) \xi
              )   )   )  +
    \xi \Phi (1,\xi )
      ( - ( \pi \xi   )  +  \nonumber\\ &&  \nonumber
       2{\zeta }^2 ( 1 + \pi \xi   )
         ( -1 + 2{\xi }^2  )  +
       {\zeta }^3 ( \pi  - 2\pi {\xi }^2
           )  + 2\zeta \xi
         ( 2 + 2\pi \xi  - 3{\xi }^2 -
          2\pi {\xi }^3  )  +
       4\zeta {\xi }^3
         ( -1 + {\xi }^2  ) \Phi (1,\xi )
        )  + \nonumber\\ \nonumber &&\zeta
      ( \zeta  - 2\xi   )
     \Phi (1,\zeta ) ( -2
         ( -1 + {\zeta }^2  ) \xi  -
       \pi  ( -1 + 2{\zeta }^2  )
         ( -1 + \xi   )
         ( 1 + \xi   )  +
       2\xi  ( -{\xi }^2 +
          {\zeta }^2 ( -1 + 2{\xi }^2  )
           ) \Phi (1,\xi )  ) \Bigg)\times\nonumber\\ \nonumber &&\Bigg({2
    { ( \zeta  - \xi   ) }^2
     ( \pi  + \xi  - \pi {\xi }^2 +
      \xi  ( -3 + 2{\xi }^2  )
       \Phi (1,\xi )  ) } \Bigg)^{-1}\nonumber -\Biggl(\frac{1}
       {2 { ( \zeta  - \xi   ) }^3} \Bigg[ ( \zeta  - \xi   )
        ( -8 + \pi  {\zeta }^3 -
         2 {\zeta }^2
           ( 1 + \pi  \xi   )  +
         \zeta  \xi   ( 10 + \pi  \xi   )
          )  - \nonumber\\ && 2
        ( {\zeta }^5 + 2 \xi  -
         3 {\zeta }^2 \xi  - 3 {\zeta }^4 \xi  +
         \zeta   ( 2 - 4 {\xi }^2  )  +
         {\zeta }^3  ( -1 + 6 {\xi }^2  )
          )  \Phi (1,\zeta ) +\nonumber\\&&
      2  ( - ( \xi
             ( -2 + {\xi }^2  )   )  +
         {\zeta }^2 \xi
           ( -2 + 3 {\xi }^2  )  +
         \zeta   ( 2 - 5 {\xi }^2 + {\xi }^4
             )   )  \Phi (1,\xi )
   \Bigg]\Biggr)^*
 \end{eqnarray}\end{widetext}
The function $B(\xi)$  is determined by an expression
 \begin{equation}\label{B}
    B(\xi)=\frac{1}{2}~\frac{\left( -1 + {\xi }^2 \right)
    \left( -\pi  + 2\xi \Phi(1,\xi) \right) }{
     \pi  + \xi  - \pi {\xi }^2 +
     \Phi(1,\xi)\xi( 2{\xi }^2-3)   },
\end{equation}

    The quantity  $\tilde{\alpha}_{xxx}$ has a static ($\omega\to 0$)
    limit:
 \begin{eqnarray}
&&\tilde{\alpha}_{xxx}(\zeta,\zeta)=-\zeta {B(\zeta )}^2\Bigg( \pi
- 8\zeta \Phi (1,\zeta ) +
    2\left( -1 + 6{\zeta }^2 \right)\times\nonumber\\ &&\Phi (2,\zeta ) + 4\zeta(1-2\zeta^2) \Phi (3,\zeta )  + 2{\zeta }^2(\zeta^2-1)\Phi (4,\zeta )
    \Bigg)\end{eqnarray}
       If additionally   $\zeta\to 0$,
    \begin{equation}\label{100}
    \tilde{\alpha}_{xxx}(\zeta,\zeta)\approx \frac{1}{6}-
  \frac{1}
   {3 \pi } \zeta \log (\zeta/2)       -\zeta\frac{ \left( 4 +
         3 {\pi }^2 \right)
        }{12 \pi }
   + ...
\end{equation}
The case of a clean sample gives the limit 1/6. The positive slope
of  the function $\tilde{\alpha}_{xxx}(\zeta,\zeta)$ change to
negative at very low numerical value of $\zeta\sim
2\exp(-1-3\pi^2/4)$. This behavior is plotted on the inset in
Fig.2.

If $\zeta\to\infty$,
\begin{equation}\label{121}
 \tilde{\alpha}_{xxx}(\zeta-i\Omega,\zeta)\approx\frac{\pi}{8\,{\zeta }^3}
 +...
\end{equation}
If the frequency  goes to infinity,
\begin{eqnarray}\label{122}  \nonumber
 &&\tilde{\alpha}_{xxx}(\zeta,\zeta+i\Omega)= \frac{1}{{\Omega }^2}
 F(\zeta),\\
&&F(\zeta)=\frac{\pi\zeta}{4},  ~~~~~~~~\mbox{if}~~~~~~~ \zeta\to 0,\\
&&F(\zeta)=\frac{\pi}{24\zeta},~~~~~~~~\mbox{if} ~~~~~~~ \zeta\to
\infty.   \nonumber
\end{eqnarray}
The function $F(\zeta)$ has the maximum equal to 0.0717 at
$\zeta=0.455$. For finite $\omega$ and  $\zeta\to 0$
$\tilde{\alpha}_{xxx}\to 0 $.

The static limit of $\tilde{\alpha}_{xyy}$ is given by
\begin{widetext}
 \begin{eqnarray} &&\tilde{\alpha}_{xyy}(\zeta,\zeta) =
-\Bigg(2 {\pi }^2 \zeta
     {\left(
         {\zeta }^2 -1\right) }^
      2 + \pi
     \left( 1 +2
       {\zeta }^4-6\zeta^2\right)+2 \left( \zeta  +
       2 {\zeta }^5-2\zeta^3 \right)+
          \Phi(1,\zeta)( 8\zeta^6 - 16\zeta^4-6\zeta^3+11\zeta^2 )
         + \nonumber\\ &&
4\zeta^3\left(2{\zeta }^4 -4\zeta^2+3   \right)\Phi^2(1,\zeta)
                     \Bigg) \Bigg[2
    \left(
      {\zeta }^2 -1\right)
    \left( \pi  + \zeta  -
      \pi  {\zeta }^2   +
       {\zeta } \Phi(1,\zeta)(2\zeta^2-
      3)
      \right) \Bigg]^{-1}. \nonumber\\\end{eqnarray}\end{widetext}
which yields
 \begin{equation}\label{103}
    \tilde{\alpha}_{xyy}(\zeta ,\zeta)\approx-\frac{1}{2} +\frac{\zeta}{2\pi}\Big(3\log\frac{\zeta}{2}+3+2\pi^2\Big)+...
       ~~~~~~~~\mbox{if}~~~~~~\zeta\to 0,
\end{equation}
The function  $\tilde{\alpha}_{xyy}(\zeta ,\zeta)$  (similarly to
$\tilde{\alpha}_{xxx}(\zeta ,\zeta)$) has singularity at $\zeta=0$
and changes the sign of slope at very low $\zeta$.

 For large  $\zeta$ we have
   \begin{equation}\label{1031}
    \tilde{\alpha}_{xyy}(\zeta+i\Omega ,\zeta)\approx -\frac{5\pi}{24\,{\zeta }^3} +...
    ~~~~~~~~\mbox{if}~~~~~~\zeta\to \infty. \end{equation}
   If $\zeta\to 0$,
    \begin{equation}\label{1032}
    \tilde{\alpha}_{xyy}(\zeta+i\Omega ,\zeta)\approx - \pi\zeta\Big(\frac{1+2\Omega^2}
    {2\Omega\sqrt{1+\Omega^2}}-1\Big). \end{equation}
The high-frequency behavior of $\tilde{\alpha}_{xyy}(\zeta+i
\Omega ,\zeta)$ is \begin{widetext}
\begin{eqnarray}
\tilde{\alpha}_{xyy}\approx\frac{\zeta^2(1-\zeta^2)}{2\Omega^2}\frac{3
\pi - 4 \zeta  -
  2 {\pi }^2 \zeta  -
  2 \pi  {\zeta }^2 +
  2 {\pi }^2 {\zeta }^3 +
  \left( 8 \pi  {\zeta }^2 -\pi  +
           4 {\zeta }^3 -
     8 \pi  {\zeta }^4
     \right)  \Phi(1,\zeta)  +
  8\zeta^3\left( {\zeta }^2-1 \right)
   {\Phi(1,\zeta) }^2}{\pi(\zeta^2-1)-\zeta+\zeta(3-2\zeta^2)\Phi(1,\zeta)}
      \end{eqnarray}\end{widetext}
with asymptotics
\begin{equation}
\tilde{\alpha}_{xyy}\approx
-\frac{\zeta^2}{2\Omega^2}(3+\log\frac{\zeta}{2})...~~~~\mbox{for}~~~\zeta\ll
1,\end{equation}and\begin{equation}
\tilde{\alpha}_{xyy}\approx\frac{\pi}{24\zeta\Omega^2}+...~~~~
\mbox{for}~~~\Omega\gg \zeta\gg 1.
\end{equation}
In the static limit the component  $\tilde{\alpha}_{yxy}$  can be
presented in the form
  \begin{widetext}\begin{eqnarray}\nonumber\alpha_{yxy}(\zeta,\zeta)=\frac{1}{{6 {\zeta }^2
    { (  {\zeta }^2 -1 ) }^3}}\Bigg(2  ( {\zeta }^2 -1 )
      (2 {\zeta }^4  -1 - {\zeta }^2 +
       3 {\zeta }^2 \Phi(1,\zeta)   )-
    3 {\zeta }^2  (  (  {\zeta }^2 -1
           )   ( 16 {\zeta }^2-11 - \\
          8 {\zeta }^4 +
          4 \pi  \zeta
           { ( {\zeta }^2 -1 ) }^2  )
        +  ( 15 {\zeta }^2 -6 -
          20 {\zeta }^4 + 8 {\zeta }^6  )
        \Phi(1,\zeta)   )  B(\zeta )\Bigg)\end{eqnarray}\end{widetext}

  For small or large values of $\zeta$ we have
 \begin{eqnarray}
 \tilde{\alpha}_{yxy}(\zeta,\zeta)\approx -\frac{1}{3\zeta^2}
 +...
 ~~~~~\mbox{if}~~~~~~ \zeta\ll 1, \nonumber\\
 \tilde{\alpha}_{yxy}(\zeta,\zeta)\approx -\frac{\pi }{3\zeta
 ^3}+...
 ~~~~~\mbox{if}~~~~~~ \zeta\gg 1.
 \end{eqnarray}
 For large $\Omega$
  \begin{eqnarray}  \label{asympReImalyxy}
  &&\mbox{Re}(\tilde{\alpha}_{yxy})\approx\frac{F_1(\zeta)}{\Omega^2},
  ~~\mbox{Im}(\tilde{\alpha}_{yxy})\approx\frac{F_2(\zeta)}{\Omega^3},\\
  && \label{F1} F_1(\zeta)=-\frac{1}{4} \Bigg(20 - 48 {\zeta }^2 + {\pi }^2
   \left( 1 - 4 {\zeta }^2 \right)  +
  6 \pi  \zeta  \left( -1 + 4 {\zeta }^2 \right) \nonumber
      \\  && +2 \left(  26 {\zeta }^2 - 4 -
     24 {\zeta }^4 + \pi  \zeta
      \left(4 {\zeta }^2 -3\right)  \right)
   \Phi(1,\zeta)\Bigg),\\ \label{F2} &&F_2(\zeta)=
   \frac{1}{24}({\pi }^2 \left( 3 \zeta  - 12 {\zeta }^3 \right)-\nonumber96
\zeta
 \left(  {\zeta }^2 -4\right) \\&&
      + 2 \pi  ( 25 - 88 {\zeta }^2 +
     24 {\zeta }^4 )\nonumber\\&&  - 2 \zeta  \left( 144 + 9 \pi  \zeta  -
     224 {\zeta }^2 - 12 \pi  {\zeta }^3 +
     48 {\zeta }^4 \right)  \Phi(1,\zeta) ).~~~~~~\end{eqnarray}

     For small and large $\zeta$ this gives
\begin{eqnarray}
&&\mbox{Re}(\tilde{\alpha}_{yxy})\approx\frac{1}{\Omega^2} (-5 -
\frac{{\pi }^2}{4}  - 2\ln (\frac{\zeta }{2})),~\mbox{if}
~~~\zeta\to 0, \nonumber \\
&&\mbox{Re}(\tilde{\alpha}_{yxy})\approx-\frac{\pi}{6\Omega^2\zeta},
~\mbox{if} ~\zeta\to \infty.
\end{eqnarray}
\begin{eqnarray}
&&\mbox{Im}(\tilde{\alpha}_{yxy})\approx\frac{1}{\Omega^3}(\frac{25\pi}{12}+
\zeta(16+\frac{\pi^2}{8}+12\ln\frac{\zeta}{2})),~~\mbox{if} ~~\zeta\to 0,\nonumber\\
&&\mbox{Im}(\tilde{\alpha}_{yxy})\approx
\frac{1}{\Omega^3}(\frac{\pi}{12}+\frac{76}{45\zeta}),~\mbox{if}
~\zeta\to \infty.
\end{eqnarray}

\section{Numerical results and discussion }

\begin{figure}[ht]
\includegraphics[width=7.5cm]{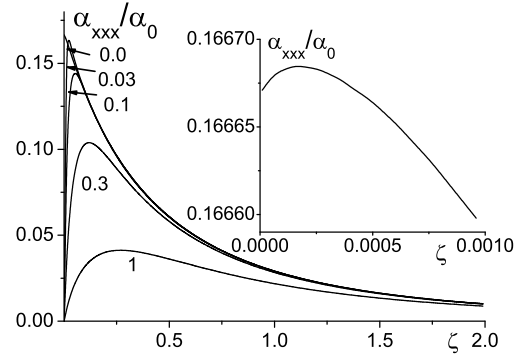}
\leavevmode\caption{The dependence of  $ \tilde{\alpha}_{xxx}$ on
the parameter $\zeta=\tau/\tau_i$ for different frequencies
$\Omega=\omega\tau=1,0.3,0.1,0.03, 0$ (marked on curves);
$\alpha_0=e^3/(4\pi^2v_FN^2D^2)$.  Insert: the dependence of  $
\tilde{\alpha}_{xxx}$ for small values of the parameter $\zeta$ at
$\Omega=0$.} \label{figalxxx}
\end{figure}

 \begin{figure}[ht]
\includegraphics[width=7.5cm]{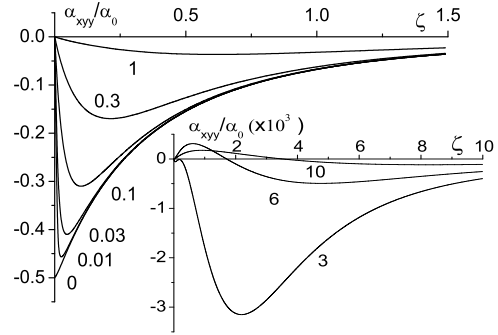}
\leavevmode\caption{The dependence of  $ \tilde{\alpha}_{xyy}$ on
$\zeta$ for the different parameters  $\Omega$  (marked on
curves). } \label{figalxyy}
\end{figure}
\begin{figure}[ht]
\includegraphics[width=7.5cm]{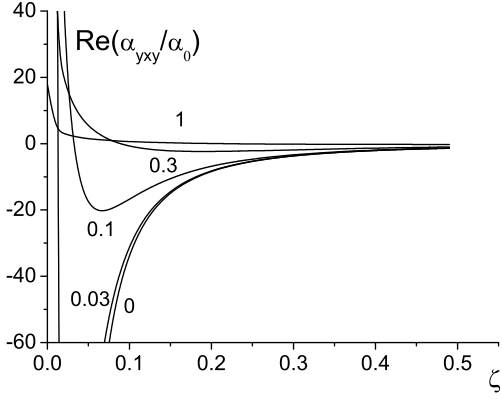}
\leavevmode\caption{The dependence of linear photogalvanic
coefficient $\mbox{Re}( \tilde{\alpha}_{yxy})$ on $\zeta$ for the
same parameters $\Omega$ as in the Figure 2.} \label{figRealyxy}
 \end{figure}
 \begin{figure}[ht]
\includegraphics[width=7.5cm]{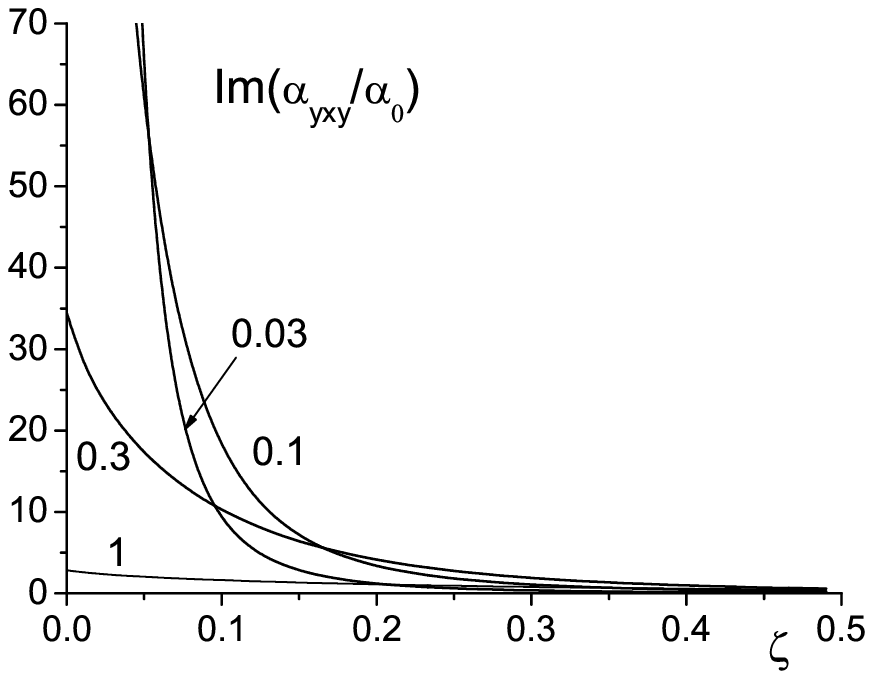}
\leavevmode\caption{The dependence of circular photogalvanic
coefficient $\mbox{Im}( \alpha_{yxy})$ for the same parameters
$\Omega$ as in the Figure 2} \label{figImalyxy}
 \end{figure}
 \begin{figure}[ht]
\includegraphics[width=7.5cm]{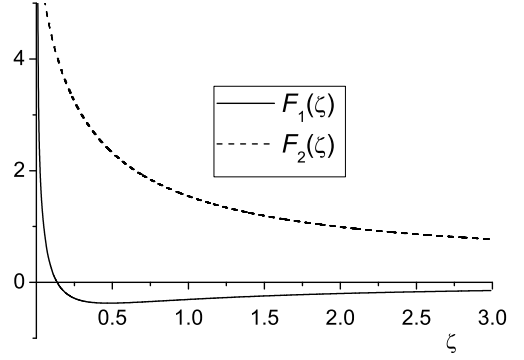}
\leavevmode\caption{Asymptotics of photogalvanic coefficient
$\alpha_{yxy}=F_1(\zeta)/\Omega^2+iF_2(\zeta)/\Omega^3$ for large
$\Omega$, according to the Equations (\ref{F1}),(\ref{F2}).}
\label{figasympt}
 \end{figure}

The Figures \ref{figalxxx}-\ref{figImalyxy} represent all
components of $ \tilde{\alpha}_{ijk}$ calculated according
Equations (\ref{alxxx}-\ref{alyxy}) at $T=0$  {\it versus}
parameter $\zeta=\tau/\tau_i$ for different frequencies. These
dependencies can be treated as the dependencies of $\alpha_{ijk}$
on the rate of impurity scattering. The sign of coefficient $
\tilde{\alpha}_{xxx}$ is positive, while the other coefficients
change sign. In accord with found asymptotics all components tend
to zero for $\zeta\to\infty$ and exhibit non-analytical behavior
at $\zeta=0$.

Asymptotic behavior $\tilde{\alpha}\propto \zeta^{-3}$ (or
 $\alpha\propto \tau^{-1}$) at large  $\zeta$ follows from the
odd dependence of the current on the asymmetric scattering on the
cuts that results in the proportionality of the current to the
scattering rate on cuts for low their concentration. This
asymptotics corresponds to case of weak asymmetric scattering
usually considered in the theory of PGE.

The value $\tilde{\alpha}_{ijk}(0,0)$ depends on the order of
limit $\omega\to 0$, $1/\tau_i\to 0$: for example, if first
$\omega\to 0$ then $1/\tau_i\to 0$ $ \tilde{\alpha}_{xxx}\to 1/6$,
else $ \tilde{\alpha}_{xxx}\to 0$. Such behavior results from the
absence of relaxation of electrons  moving along $y$ axis. The
state with $p_x=0$ plays role of a drain for electrons. These
electrons do not participate in the transport along $x$ axis, but
due to absence of relaxation they accumulate in the state $p_x=0$;
this suppresses the distribution function with finite $p_x$, and
$j_x \to 0$. On the contrary, the transport along $y$ axis
diverges due to the same reason.

For linear polarized electric field the signs of  current
components depend on the direction of polarization. Physically,
this can be explained by the effective increase of the
mean-squared component of electron momentum along field and
subsequent increase (decrease) of scattering on the cuts.  Let
electron with a momentum ${\bf p}=(\pm p,0)$ impacts with a cut.
The change of momentum are  equal to  $-2p$ for an electron with
the momentum ${\bf p}=(p,0)$ and $(1+2/\pi)p$ for an electron with
the opposite momentum, respectively. In equilibrium this change is
compensated by the contributions of other electrons. But the
increase of the mean-squared component of electron momentum along
field gives the finite positive contribution to the total current.
The acceleration of an electron by the field in $y$ direction
increases  $y$ component of  the momentum and produces the
opposite direction of $j_x$.

The values of coefficients $\alpha_{yxy}$ are essentially larger
than  $\alpha_{xxx}$ and $\alpha_{xyy}$. It is a consequence of
the fact that the motion along $y$ direction is collisionless
unless the impurity scattering is taken into account. Obviously,
the difference is more pronounced at low $\zeta$.

It is useful to calculate the possible maximal PGE coefficient.
From said above it follows that this maximum is achieved for the
component $Re(\alpha_{yxy})$ at low frequencies $\omega\ll
1/\tau_i$ and for clean material $\tau_i\gg\tau$. In this case we
have
\begin{equation}
  j_y=-\frac{1}{48\pi}n_eev_F \left(\frac{eEl_i}{\epsilon_F}\right)^2\sin
  2\theta,
\end{equation}
 where  $n_e$ is the electron concentration, $l_i=v_F\tau_i$,   $\theta$ is the
 angle  of electric field with respect to the $x$ direction. The
 estimation gives the value $j_y\sim 10^{-5}$A/cm for  $l_i=10^{-3}
 cm$, $n_i=10^{12}\mbox{cm}^{-2}$, $v_F\sim 10^7$cm/s.

The special case  is the  limit of zero impurity concentration. As
we have found, in this case for finite frequency the current in
$x$ direction becomes equal to zero and infinity for $y$
direction. From said above it is evident that the inelastic
collisions become essential: they limit this non-analytical
behavior. The careful examination of this question is beyond the
paper.

The special case  is the zero-frequency limit. As we have found,
in this case  for finite impurity scattering the current in $x$
direction becomes equal to zero and infinity for $y$ direction.
From said above it is evident that the inelastic collisions become
essential: they limit this non-analytical behavior. The careful
examination of this question is beyond the paper.

We have restricted ourselves by the classical consideration only.
Two quantum factors have not been taken into account:
comparability of $\hbar \omega$ with characteristic electron
energies,  the Fermi energy and the temperature, and the quantum
corrections to conductivity. It should be emphasized that in used
approximation the PGE in the metal case does not depend on the
temperature in the low temperature limit. This is because the
momentum relaxation in metal is temperature-independent (unlike
energy relaxation). Despite the involvement of energy relaxation
into the control on the distribution function, the current in the
second order in electric field is determined by the contributions
which do not depend on energy relaxation.

The quantum corrections suppress the low temperature transport and
should lead to the temperature dependence of the effect.  Both
modelling of \cite{semidisk1,cristadoro}
 and the present theory   neglect
quantum corrections.   Thereupon, the temperature dependence of
steady current in  \cite{semidisk1,cristadoro} is not clear.  The
approach of these papers   used a
 friction force essentially  depending on the excess electron
energy. Possibly, it is the reason of the temperature dependence.
 It should be mentioned that the quantum
corrections are of less importance in high-mobility samples
utilized usually in experiments with antidots.

\section{Comparison with numerical simulations}
 In this section we compare our results with the  simulations
 of the  flow velocity of a particle colliding with semidisks.\cite{semidisk2} We
 consider the model used here as a simplification of the
 semidisks model of Ref. \cite{semidisk2}. The essential differences between approaches
  of Ref.\cite{semidisk2} and the present paper are: periodic/random distribution of
 asymmetric scatterers, deterministic/chaotic character of
 electron scattering on antidots and
 motion between them,  and absence/presence of the gas approximation.
 However,  the case of low  density of semidisks and weak
 friction corresponds to the gas approximation because the scattering randomizes the
 motion well enough, and the results of both approaches should be in
 accord with each other.

 The Fig.6 in \cite{semidisk2} shows approximate quadratic
 dependence of the flow velocity on the alternating field what
 agrees with the second-order in field approximation used in the
 present paper.

 The dependence Fig.7 from \cite{semidisk2} demonstrates the drop of the
 flow velocity with the growth of the semidisk size.  This
 behavior qualitatively corresponds to the drop $\alpha_{xxx}\to
 0$ for $\tau\to 0$ (Fig.2 of present paper). At the
 same time there is no drop in this figure for large $\tau\to\infty$
 following from the present paper.

 The dependence of the flow velocity on the distance between
 semidisk centers (Fig.8 from \cite{semidisk2}) can be treated
 as rescaled   dependence of   $\alpha_{xxx}$ on the mean free
 time $\tau$ which in the present case vanishes both for $\tau\to
 0$ and  $\tau\to\infty$ (Fig.2 of the present paper).

 The dependence of  the flow velocity (Fig.9 in \cite{semidisk2}) on the impurity scattering
 also has a drop for  $\tau_i\to 0$, like expected from the
 Eq.(\ref{121}), but
 there is no  drop in this figure for large $\tau_i$.

 Hence, there is a partial qualitative accordance between the present
 results and computer simulations of \cite{semidisk2}. The origin
 of discrepancy needs additional study.

\section{Conclusions}
The considered system of oriented asymmetric scatterers-cuts has
$\alpha_{xxx}$, $\alpha_{xyy}$, and $\alpha_{yxy}$ non-zero
components of photogalvanic tensor. The linear photogalvanic
effect is determined by $\alpha_{xxx}$, $\alpha_{xyy}$, and
$\mbox{Re}(\alpha_{yxy})$. The circular-polarized illumination,
causes the response of the $y$ component of current only,
determined by $\mbox{Im}(\alpha_{yxy})$. The static limit of the
current in impurity-free system is ambiguous, depending on value
of the product of frequency to the impurity mean-free time.   The
$x$ component of the current  is limited and in the impurity-free
system tends to zero, while the $y$-component tends to infinity.
This is explained by the accumulation of electrons in the state
with zero $x$-component of momentum.

 The authors are grateful to D.L. Shepelyansky and A.D. Chepelyansky
 for numerous helpful  discussions of
  the problem.  The work was supported by grants of RFBR Nos. 05-02-16939 and
04-02-16398, Program for support of scientific schools of the
Russian Federation No. 593.2003.2 and INTAS No. 03-51-6453.

\end{document}